\documentclass[aps,pre,amsmath,amssymb,twocolumn,floatfix,showpacs]{revtex4}
\usepackage{graphicx}
\usepackage{epsfig}
\newcounter{abc}

\input epsf

\newcommand{\be}{\begin{equation}}
\newcommand{\ee}{\end{equation}}
\newcommand{\bea}{\begin{eqnarray}}
\newcommand{\eea}{\end{eqnarray}}

\newcommand{\f}{f}

\newcommand{\p}{^{(0)}}

\newcommand{\one}{^{(1)}}
\newcommand{\two}{^{(2)}}

\newcommand{\ronesig}{\rho\one(\sigma)}
\newcommand{\rtwosig}{\rho\two(\sigma)}

\newcommand{\rset}{\{\vec{r}\}}

\newcommand{\uset}{\{\vec{u}\}}

\newcommand{\RsetCS}{\{ \vec{R}\}^{CS}}

\newcommand{\RsetF}{\{ \vec{R}\}^{F}}

\newcommand{\Rsetgamma}{\{ \vec{R}\}^{\gamma}}
\newcommand{\Rsetgammaprime}{\{ \vec{R}\}^{\gamma^{\prime}}}

\newcommand{\zratio}{{\mathcal{R}}_{\mbox{\sc {f,cs}}}}

\usepackage{graphicx}
\usepackage{bm}
\usepackage{epsfig}

\begin{document}

\title{Solid-liquid coexistence of polydisperse fluids via simulation}

\author{Nigel B. Wilding}
\affiliation{Department of Physics, University of Bath, Bath BA2 7AY, U.K.}

\begin{abstract} 

We  describe a simulation method for the accurate study of the
equilibrium freezing properties of polydisperse fluids under the
experimentally relevant condition of fixed polydispersity. The approach
is based on the phase switch Monte Carlo method of Wilding and Bruce
[Phys. Rev. Lett. {\bf 85}, 5138 (2000)]. This we have generalized to
deal with particle size polydispersity by incorporating updates which
alter the diameter $\sigma$ of a particle, under the control of a
distribution of chemical potential differences $\tilde\mu(\sigma)$.
Within the resulting isobaric semi-grand canonical ensemble, we detail
how to adapt $\tilde\mu(\sigma)$ and the applied pressure such as to
study coexistence, whilst ensuring that the ensemble averaged density
distribution $\rho(\sigma)$ matches a fixed functional form. Results are
presented for the effects of small degrees of polydispersity on the
solid-liquid transition of soft spheres.

\end{abstract} 
\maketitle 
\epsfclipon  

\section{Introduction and background}

\label{sec:intro} 

Our ability to accurately determine the phase coexistence
properties of model systems via computer simulation has increased
greatly in recent years. Progress has come in the form of specialized
Monte Carlo methods that are tailor-made for tackling the core problem
of coexistence, namely the comparison of the statistical weights
of the disjoint regions of configuration space associated
with competing phases. To achieve this one generally aims to engineer a
sampling path which {\em connects} the phases in question, allowing each
to be visited repeatedly in the course of a single simulation run. The
reward for doing so comes in the form of direct, accurate and
transparent measurements of free energy differences and coexistence
parameters \cite{Bruce2003}. 

In practice, however, the most obvious choice of an inter-phase sampling path,
namely one routed via mixed-phase (ie. interfacial) configurations may
not always be the most straightforward to negotiate. Typically 
such regions of configuration space are characterized by large free
energy barriers, or by a `pitted' free energy landscape  -- features
which hinder efficient sampling.  Contemporary strategies seek to either
surmount or circumvent these impediments (for a review see
\cite{Bruce2003}). One such scheme --  which is in principle rather
general -- is phase switch Monte Carlo (PSMC).  Its key feature is a
phase space `leap' \cite{bruce2000} that directly maps a
pure phase configuration of one phase onto a pure phase configuration of
another. As such it completely avoids mixed-phase configurations
and any attendant sampling difficulties.

In the context of fluid-solid coexistence, PSMC permits efficient
two-phase sampling and hence the direct determination of equilibrium
freezing parameters and their uncertainties. This contrasts with
simulation approaches which seek to emulate physical crystal nucleation
processes, which commonly encounter significant problems, principally a
large degree of metastability of the phases, extended timescales for
crystallization and a tendency for the crystals formed to exhibit
vacancies, interstitials and other defects. To date, PSMC has been
successfully applied to the freezing of hard spheres, softly repulsive
spheres, and the Lennard-Jones fluid
\cite{wilding2000,errington2004,mcneil-watson2006,Wilding2008}.

The present paper is concerned with extending the applicability of PSMC
to the study of the freezing properties of model {\em polydisperse}
fluids. Polydispersity is the feature, pervasive in soft condensed matter
systems such as colloids and liquid crystals, whereby the constituent
particles are not all identical but instead exhibit variation in terms
of some attribute ($\sigma$, say) such as the particle size or shape, etc. 
It leads to phase behaviour that is considerably richer in
both variety and form than that of monodisperse systems \cite{Sollich2002}.

Typically one describes the polydispersity of a given system in terms of
a density distribution $\rho\p(\sigma)$ which counts the number of
particles per unit volume having $\sigma$ in the range
$\sigma\dots\sigma+d\sigma$. In most real polydisperse systems the form
of $\rho\p(\sigma)$ is {\em fixed} by the synthesis of the fluid, and
only its scale can change according to the degree of dilution of the
system. Accordingly, one writes $\rho\p(\sigma)=n\p f(\sigma)$
where $f(\sigma)$ is a normalized fixed shape function and $n\p$ is
the overall number density. Varying $n\p$ corresponds to scanning a
``dilution line'' of the system.

At coexistence, the particles of the various $\sigma$
values will be partitioned unequally between the phases. This is the
phenomenon of ``fractionation'' which underpins the opulence of
polydisperse phase behaviour.  To describe fractionation it is necessary
to define separate ``daughter'' distributions $\rho^{(i)}(\sigma)$
(\:$i=1,2...$) which measure the distribution of the polydisperse
attribute for each phase $i$. When the polydispersity is fixed,
conservation of particles implies that the weighted average of the
daughter distributions equals the fixed overall density distribution,
or ``parent'' $\rho\p(\sigma)$, For instance, at two-phase coexistence
$\rho\p(\sigma)=n\p
f(\sigma)=(1-\xi)\rho^{(1)}(\sigma)+\xi\rho^{(2)}(\sigma)$, with $1-\xi$
and $\xi$ the respective fractional volumes of the phases. This
expression represents a generalisation of the Lever rule to polydisperse
systems.

To illustrate the central role of fractionation in polydisperse phase
behaviour, it is constructive to consider first the familiar case of the
binodal curve of a monodisperse system in the density-temperature plane.
This curve serves a dual purpose: on the one hand it describes the range
of overall densities for which phase coexistence occurs; and on the
other hand it identifies the densities of the coexisting phases
themselves. Now, for a polydisperse system the range of overall (parent)
densities that leads to coexistence is similarly delineated by a
curve in $n\p$-$T$ space -- the so-called ``cloud'' curve. However, the
densities of the coexisting phases themselves do not in general coincide
with the cloud curve. Instead, as one varies the parent density $n\p$
through the coexistence region at a fixed temperature (say), one
generates an infinite sequence of pairs of differently fractionated
coexisting phases. It is customary to single out the end points of this
sequence for special attention, ie. the case of incipient phase
separation that occurs when the value of $n\p$ coincides with the cloud
curve. Under these conditions, one daughter phase has a fractional
volume of essentially unity and consequently (from the Lever rule) a
density distribution that is identical to the parent, while the other
phase -- known as the ``shadow'' -- has an infinitesimal fractional
volume and a density distribution that deviates maximally from that of
the parent. The curve formed by plotting the number density (or packing
fraction) of the shadow phase as a function of temperature is known as the
shadow curve.

Qualitative differences between the phase behaviour of monodisperse and
polydisperse systems are also evident in other projections of the full
phase diagram. For instance, in the pressure-temperature plane, 
coexistence for a monodisperse system occurs along a simple line. By
contrast for a polydisperse system, coexistence occurs within a {\em
region} of the pressure-temperature phase diagram
\cite{Rascon2003,bellier-Castella2000}. Accordingly the task of
exploring the full coexistence behaviour of a polydisperse system
represents a far greater endeavor than for a monodisperse system.

Owing to the computational complexities associated with  polydispersity,
simulation studies of its effects on fluid-solid phase equilibria have
been rather sparse. In pioneering work, Bolhuis and Kofke employed an
isobaric semi-grand canonical ensemble approach to study the effects of
particle size polydispersity on hard sphere freezing
\cite{bolhuis1996b,kofke1999}. Within this ensemble, Monte Carlo updates
are performed that allow particles to change diameter, under the control
of an imposed distribution of chemical potential differences
$\tilde\mu(\sigma|\sigma_0)=\mu(\sigma)-\mu(\sigma_0)$, with $\sigma_0$
a reference particle diameter. The advantage of this approach is two
fold: it allows the instantaneous form of the density distribution
$\rho(\sigma)$ to fluctuate, thereby sampling many different
realizations of the disorder in the course of a run, and it permits
fractionation of the coexisting phases. However, in this early work no
attempt was made to {\em adapt} the form of $\tilde\mu(\sigma)$ in order
to ensure that the ensemble averaged density distribution
$\bar\rho(\sigma)$ had a fixed functional form. Instead the activity
distribution $\exp[\beta\tilde\mu(\sigma)]$ was assigned a Gaussian
form, peaked at $\sigma_0$, and various widths of the Gaussian were
studied in order to change the degree of polydispersity. In such a fixed
chemical potential representation, $\bar\rho(\sigma)$ can vary
dramatically across the phase diagram. Additionally, coexistence occurs
only along a line in the $p-T$ plane (as in the monodisperse case),
rather than within a region as occurs experimentally. These artifacts
limit the applicability of the results.

To obtain the properties of the fluid-solid phase boundary, Bolhuis and
Kofke employed Gibbs-Duhem integration (GDI) techniques. This approach
traces a coexistence line in pressure-temperature space by integrating
its derivatives, obtainable from measurements of single phase properties
near coexistence. The principal strengths of GDI is that it can quickly yield
an estimate for a phase boundary and avoids mixed phases configurations.
However, independent prior knowledge of a coexistence state
point is required in order to bootstrap the integration, and since there is no
reconnection of the two phases beyond the initial starting point,
integration errors can accumulate. The magnitude of these errors is
difficult to quantify because there is no feedback mechanism to indicate
when the integration has departed from the true coexistence line
(provided one remains within the rather wide band of metastable states
that borders this line).     

In more recent simulation work, Fern\'{a}ndez {\em et al}
\cite{Fernandez2007} used a canonical ensemble approach to investigate
the freezing of polydisperse soft spheres. However, owing to the small
system size employed, the physical separation of the phases that
would normally be expected to occur in a canonical ensemble simulation was
(for the most part) absent and hence fractionation of the individual
phases could not be measured. Consequently the ``phase boundary''
presented by Fern\'{a}ndez {\em et al} contains no information on the
properties of any of the infinity of coexisting fluid-solid pairs,
serving at best only as a rough estimate of the middle of the range of parent
densities for which coexistence might occur.

The method described in the present paper permits the accurate study of
fluid-solid coexistence for systems of particles whose polydispersity is
fixed. Accuracy is obtained by embedding the isobaric semi-grand
canonical ensemble within the PSMC framework, thus allowing the direct
connection of the coexisting phases. Fixed polydispersity is realized by
implementing an iterative reweighting scheme that adapts
$\tilde\mu(\sigma)$ and the pressure $p$ such that the conjugate density
distribution $\bar\rho(\sigma)$ matches a prescribed parent form, even
within the coexistence region. Our paper is arranged as follows. In
sec.~\ref{sec:psmc} we provide a summary of principal aspects of PSMC as
applied to monodisperse systems, before proceeding to detail the
extensions necessary to deal with polydispersity. Thereafter we describe
how one can use the method to accurately determine fluid-solid
coexistence properties for systems having fixed polydispersity. Finally
we provide some illustrative results for a system of polydisperse soft
spheres, obtaining cloud points, demonstrating the broadening of the
coexistence region in the pressure-temperature plane, and quantifying
fractionation effects.

\section{Phase switch Monte Carlo}

\label{sec:psmc} 

A full description of the statistical mechanical background to PSMC as
well as an in-depth discussion of its implementation for problems of
fluid-solid coexistence has recently been given elsewhere (see
ref.~\cite{mcneil-watson2006}) and shall not therefore be repeated here
in full. Instead we shall focus somewhat more on the generalities of the
method, as a prelude to describing the extensions required to deal with
polydispersity.

\subsection{Monodisperse systems}
\label{sec:mono}

Let us work within an isobaric-isothermal (constant-{\em NpT}) ensemble, in
which the particle number $N$, pressure $p$ and temperature $T$ are all
prescribed \cite{FrenkelSmit2002}. Then the relative stability of fluid (F) and crystalline solid
(CS) phases is determined by the difference in their Gibbs free
energies.  This can in turn be obtained from the ratio of the {\em
a-priori} probabilities of the phases
\cite{Bruce2003,mcneil-watson2006}:

\begin{equation} 
\label{eq:gibbs} 
\Delta g \equiv g_{CS}(N,p, T)-g_{F} (N,p,T)  \equiv \frac{1}{N} \ln {\mathcal{R}}_{\mbox{\sc {f,cs}}}\:.
\end{equation}
with

\begin{equation}
\label{eq:Prat} 
{\mathcal{R}}_{\mbox{\sc {f,cs}}} \equiv \frac{P(F| N,p,T)}{P(CS|N,p,T)} \:.
\end{equation}

In order to directly measure ${\mathcal{R}}_{\mbox{\sc {f,cs}}}$, a MC
procedure is required that samples both the solid and the fluid regions
of configuration space in the course of a single simulation run at the
prescribed values of $N,p$ and $T$. From this one simply measures the
probabilities $P(F)$ and $P(CS)$ of finding the system in each of the
respective phases, and thence the probability ratio. Below we describe
how PSMC facilitates such a measurement.
  
PSMC takes as its starting point the specification of a
reference configuration $\Rsetgamma$ for each of the phases
(labeled $\gamma$) coexisting at the phase boundary.  The specific
choice of $\Rsetgamma$ is arbitrary, the only condition being
that it should be a member of the set of pure phase configurations
identifiable as ``belonging'' to phase $\gamma$.  For a crystalline
phase, a suitably simple choice of $\{\vec{R}\}^{CS}$ is the set of
lattice sites; for a fluid, a suitable choice is a randomly chosen
fluid configuration.

\begin{figure}[h]
\includegraphics[type=eps,ext=.eps,read=.eps,width=9.0cm,clip=true]{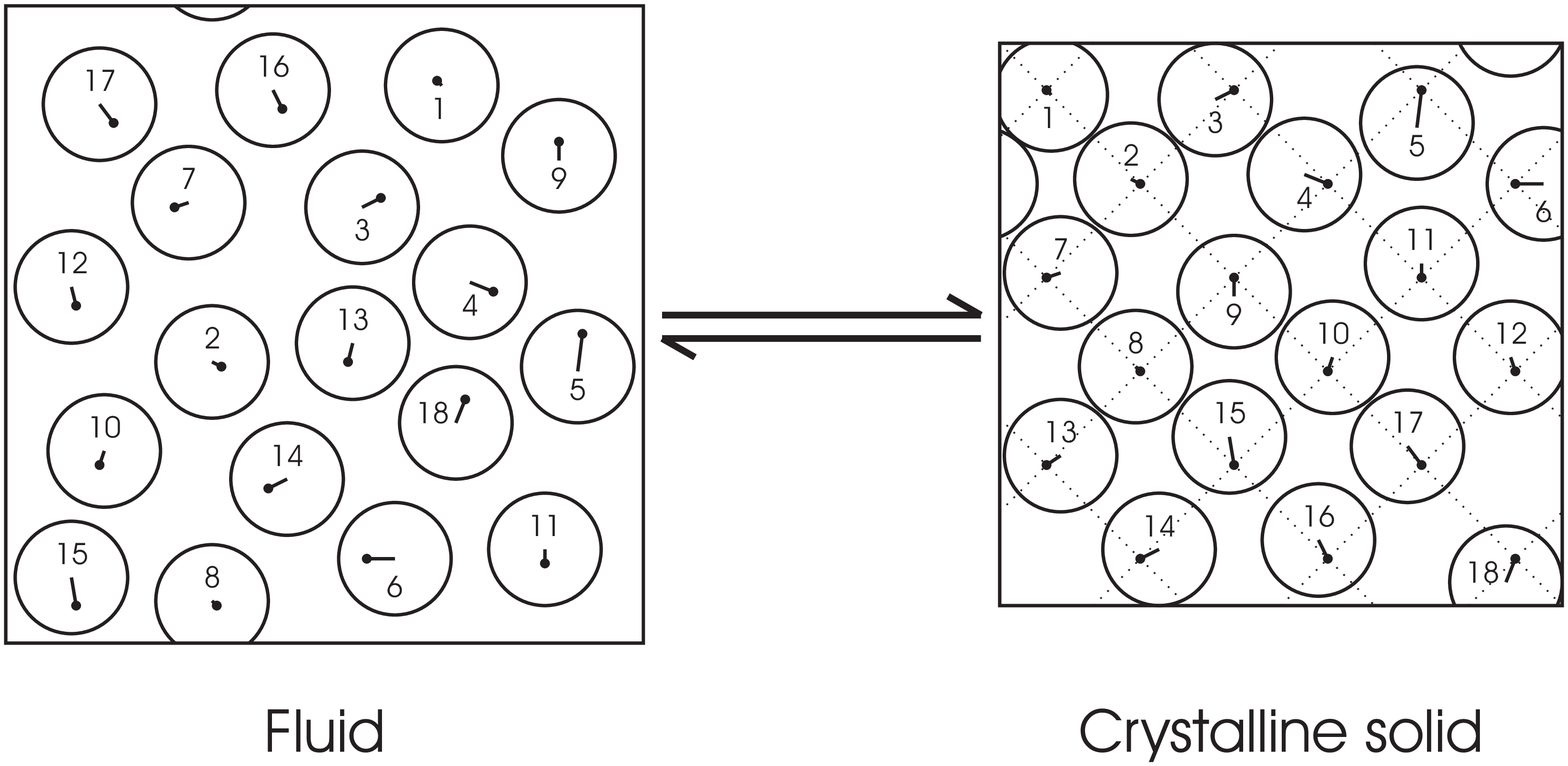}
\caption{Schematic illustration of the phase switch mechanism. The dots identify
the representative sites $\Rsetgamma$ in each phase; the displacement
vectors $\uset$ connect the centers of the distinguishable (numbered)
particles to these sites. The switch
operation shown swaps the representative sites of one phase for those
of the other phase, whilst maintaining $\uset$ constant. The particular
configuration $\uset$ shown is a ``gateway'' state (see text) because the
magnitude of the effective energy change under the switch is small.}
\label{fig:psschem}
\end{figure}

The next step is to express the coordinates of each particle in phase
$\gamma$ in terms of the displacement from its reference site, i.e. 

\begin{equation}
\vec{r}^\gamma_i=\vec{R}^\gamma_i+\vec{u}_i\:.
\end{equation}
Now, for displacement vectors that are sufficiently small in magnitude,
one can clearly reversibly map any configuration 
$\{\vec{r}^\gamma\}$ of phase $\gamma$ onto a configuration of another
phase $\gamma^\prime$ simply by {\em switching} the set of reference
sites $\Rsetgamma\to \Rsetgammaprime$, while
holding the set of displacements $\{\vec{u}\}$ {\em constant}. This
switch, which forms the heart of the method, can be incorporated in a
global MC move (see fig.~\ref{fig:psschem}).

\begin{figure}[h]
\includegraphics[type=eps,ext=.eps,read=.eps,width=8.0cm,clip=true]{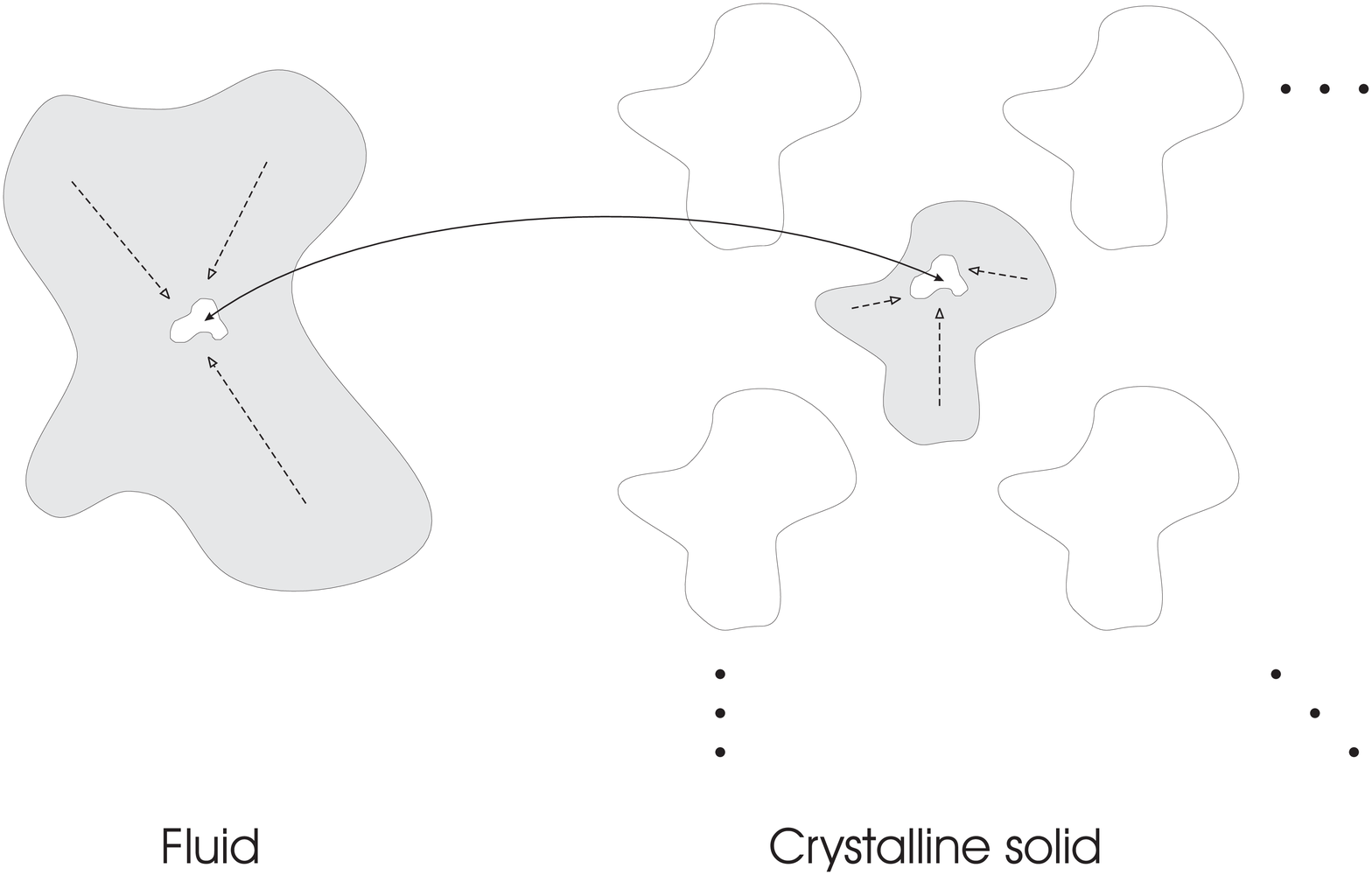}\\
\caption{Schematic illustration of the phase switch operation in terms of the
regions of configuration space associated with the fluid and crystalline
solid phases. A bias (dashed arrows) is constructed such as to enhance the probability of
the subsets of ``gateway'' states (the white islands) within the single-phase
regions, from which the switch operation (the large arrow) will be
accepted. Note that the switch accesses only one of the crystalline
phase space replica fragments associated with permutations of particles
amongst lattice sites \protect\cite{mcneil-watson2006}.}
\label{fig:blobs}
\end{figure}

A complication arises however, because the displacements $\{\vec{u}\}$
typical for phase $\gamma$ will not, in general, be typical for phase
$\gamma^\prime$. Thus the switch operation will mainly propose high
energy configurations of phase $\gamma^\prime$ which are unlikely to be
accepted as a Metropolis update. This problem can be circumvented by
employing extended sampling (biasing) techniques to seek out those
displacements $\{\vec{u}\}$ for which the switch operation {\em is}
energetically favorable. These are the gateway configurations, which
typically correspond to displacement vectors which are small in magnitude.
Fig.~\ref{fig:blobs} shows a schematic representation of the procedure.

The requisite bias is administered with respect to an ``order
parameter'' $M$. This macrovariable is defined such that the set
of typical microstates (particle configurations) associated with its
range, form a continuous phase space path linking the configurations of high
statistical weight to the gateway configurations. In our formulation, the
order parameter comes in two parts (or modes): `tether' and `energy'.
The tether mode serves to draw particles close to the representative
sites to which they are nominally associated. Then, once all particles
are within a prescribed distance of these sites, tether mode switches
off and an energy mode order parameter takes over to guide particle to
the gateway states for which the phase switch energy cost is
sufficiently small to be accepted.

To elaborate, let $M_{m,\gamma}$ denote the order parameter in mode $m$
and phase $\gamma$. Then for tether mode we write $m={\cal T}$ and
define an associated order parameter

\begin{equation}
\label{eq:tether}
M_{{\cal T},\gamma}(\uset)= \left( \frac{1}{N}\sum_{i=1}^N {\rm max}\{0,\tilde{u}_i-\tilde{u_c}\} \right)^{1/2}\:,
\end{equation}
where $\tilde{u}_i=|{\bf u}_i|/V^{1/3}$ is the distance of particle $i$
from its lattice site measured in units of the box length, and
$\tilde{u_c}$ is a prescribed dimensionless threshold radius. Only
particles whose displacement $\tilde{u}_i$ exceeds this threshold
contribute to $M_{{\cal T},\gamma}$.

The tether mode is active iff $\tilde{u}_i>\tilde{u_c}\:$ for at least one particle
$i$, i.e.~when $M_{{\cal T},\gamma}>0$. Otherwise control hands over to the
`energy' mode $m={\cal E}$; its associated order parameter is defined by

\begin{equation}
M_{{\cal E},\gamma}(\uset)=\mathrm{sgn}(\Delta {\cal E}_{\gamma^\prime\gamma})\ln(1+|\Delta {\cal E}_{\gamma^\prime\gamma}|)\:,
\label{eq:Emode}
\end{equation}
where 

\begin{equation}
\label{eq:cost}
\Delta {\cal E}_{\gamma^\prime\gamma}=({\cal E}_\gamma-{\cal E}^{\rm ref}_\gamma)-  ({\cal E}_{\gamma'}-{\cal E}^{\rm ref}_{\gamma'}) 
\end{equation}
measures the change (under the phase switch operation) of the
enthalpy ${\cal E}_\gamma(\uset,V)=\Phi_\gamma(\uset)+pV$ with
respect to its value ${\cal E}^{\rm ref}$ in the 
representative microstate $\Rsetgamma$, with the latter scaled to the
instantaneous volume of phase $\gamma$ \cite{Jackson2002,errington2004}. The
presence of the logarithm in eq.~\ref{eq:Emode} is designed to moderate
the scale of the contribution of the energy cost to $M_{{\cal
E},\gamma}$, which might otherwise become excessively large for
particles with a strongly repulsive core to their interaction potential.

Having defined a suitable order parameter, a biasing (weight) function
is constructed to allow the system to reach the gateway states and hence
-- in the course of a sufficiently long run -- switch back and forth
repeatedly from fluid to solid. This weight function comes in four
parts, one for each combination of phase and mode, and the interface
between energy and tether parts are joined for continuity at their
boundary. A number of techniques exist for constructing this weight
function, but of those we have tested, we have found transition matrix
Monte Carlo \cite{Smith1995} to be (by far) the most efficient. Details
of how to implement the transition matrix approach to obtain the weight
function in the context of PSMC have previously been described elsewhere
\cite{errington2004,mcneil-watson2006} and we refer the interested
reader to those articles for details.

Turning now to the Monte Carlo procedure for sampling the order
parameter distribution, in the context of monodisperse systems this
involves four types of update, as outlined below. 

\begin{enumerate}

\item {\em `Particle translations'}. A site (identified by one of the
vectors in $\RsetCS$ or $\RsetF$) is selected at random and a candidate
state is chosen by incrementing the position coordinate of the particle
associated with that site by a random vector whose components are drawn
from a zero-mean uniform distribution of prescribed width. This
operation changes both $\rset$ and $\uset$

\item {\em `Association swaps'}. In this operation we choose two distinct sites
at random ($i$ and $j$ say) and identify the corresponding displacement
vectors $\vec{u}_i$ and $\vec{u}_j$.  The candidate state is defined by
the replacements

\begin{eqnarray}
\vec{u}_i &\rightarrow& \vec{u}_j + \vec{R}_j - \vec{R}_i \\
\vec{u}_j &\rightarrow& \vec{u}_i + \vec{R}_i - \vec{R}_j
\end{eqnarray}

This process can be regarded as a change of association: the particle
which was associated with site $j$ is now associated with site $i$ (and
{\it vice versa}).  It changes the {\em representation} of the
configuration (the coordinates $\uset$); but it leaves the physical
configuration invariant. It need be applied only to the fluid phase
where particles can wander far from their representative sites and need
to be reined back in order to reach the gateway states.

\item {\em `Volume moves'}. The volume is also updated in the
conventional way \cite{FrenkelSmit2002}, by a random walk of prescribed maximum step size,
with particle position coordinates $\{\bf{r}\}$ and representative sites $\{
\vec{R}\}^{\gamma}$ rescaled.

\item `{\em Inter-phase switch}'. The final type of MC update is the
phase switch, which entails replacing one set of the representative
configuration vectors, $\Rsetgamma$  say, by the other,
$\Rsetgammaprime$. The switch should also incorporate an appropriate
{\em volume scaling} of the system.

\end{enumerate}

Details of the appropriate biased acceptance probabilities for these moves can
be found in the original articles \cite{wilding2000,mcneil-watson2006}.

Using these MC updates, together with a suitable estimate of the weight
function, the sampling process can be initiated. During this process one
accumulates (initially in the form of a list \cite{wilding2001}) the
joint probability distribution of the order parameter, the
volume, the energy and the phase label $\gamma$. Subsequently the effects of the imposed
bias are unfolded from this distribution (in the standard manner
\cite{Bruce2003,wilding2000,mcneil-watson2006}) to provide a direct
measure of the relative probabilities of the two phases:

\begin{equation}
\zratio =\frac {\int dM\;dV\;d{\cal E}   \;P(M,V,{\cal E},F)}  {\int dM\;dV\;d{\cal E}\;  P(M,V,{\cal E},CS)} \:,
\end{equation}
from which the Gibbs free-energy-density difference follows directly via
Eqs.~\ref{eq:gibbs} and \ref{eq:Prat}. In practice these integrals are
estimated by simple sums over the list of sampled values of $M,V,{\cal
E},\gamma$. Use of histogram reweighting (HR) \cite{ferrenberg1989}
permits exploration of values of the $p$ and $T$ in the neighbourhood of
a given simulation state point and serves as an invaluable aid for
pinpointing the coexistence parameters at which the free energy
difference vanishes.

\subsection{Extension to polydisperse systems}

The natural choice of simulation ensemble for the study of $F$-$CS$
coexistence in polydisperse systems is the isobaric,
semi-grand-canonical ensemble \cite{Kofke1988}. Within this ensemble,
the particle number $N$, pressure $p$, temperature $T$, {\em and} a
distribution of chemical potential differences
$\tilde\mu(\sigma|\sigma_0)$ are all prescribed, while the system volume
$V$, the energy, and the form of the instantaneous density distribution
$\rho(\sigma)$ all fluctuate.  The fluctuation in $\rho(\sigma)$,
although subject to the constraint $V\int\rho(\sigma)d\sigma=N$,
nevertheless permits the sampling of many realizations of the
polydisperse disorder, thus ameliorating finite-size effects. Moreover, in
conjunction with volume fluctuations,  it facilitates separation into
differently fractionated phases.
 
Operationally, the sole difference between the isobaric,
semi-grand-canonical ensemble and the constant-{\em NpT} ensemble used for
PSMC is that one implements MC updates that select a particle at random
and attempt to change its diameter $\sigma$ by a random amount drawn
from a zero-mean uniform distribution. This proposal is accepted or
rejected with a Metropolis probability controlled by the change in the
internal energy and chemical potential \cite{Kofke1988}:

\[
p_{\rm acc}={\rm min}\left[1,\exp{(-\beta[\Delta \Phi+\tilde\mu(\sigma)-\tilde\mu(\sigma^\prime)])}\right]\:,
\]
where $\Delta \Phi$ is the internal energy change associated with the
resizing operation. Accordingly, one can readily extend the PSMC framework to deal with polydispersity by
specifying a form for $\tilde\mu(\sigma|\sigma_0)$ and supplementing the four
Monte Carlo operations described in Sec.~\ref{sec:mono} with a resizing move.
For the purposes of histogram reweighting, one samples the joint
probability distribution of the order parameter, the volume, the energy
and the instantaneous density distribution $\rho(\sigma)$. Again this
can be conveniently stored as a simple list of sampled values.

\section{Fixed polydispersity}
\label{sec:iterate}

For simulations of fixed polydispersity at some given $N$ and $T$, one
requires that both the pressure $p$ and $\tilde\mu(\sigma)$ are such
that a suitably defined ensemble-averaged density distribution matches
the prescribed parent $\rho\p(\sigma)=n\p f(\sigma)$. Unfortunately, the
task of determining the requisite $p$ and $\tilde\mu(\sigma)$ is
severely complicated by the fact they are an unknown {\em functional} of
the parent. Essentially, therefore, one is faced with solving an inverse
problem \cite{wilding2003a}.

To deal with this problem we employ a scheme similar to one recently
proposed in the context of grand canonical ensemble studies of
polydisperse phase coexistence \cite{buzzacchi2006}.  The method relies
on the fact that data accumulated at a given set of $p$ and
$\tilde\mu(\sigma)$ can be used to extrapolate (via histogram
reweighting) to other nearby sets of $p$ and $\tilde\mu(\sigma)$, {\em
without} further simulation. In our description we shall  assume for
convenience that while the particle size $\sigma$ is to be treated as a
continuous variable, distributions defined on $\sigma$, such as
$\rho(\sigma)$, and $\tilde\mu(\sigma)$, are represented as histograms
formed by discretising the range of $\sigma$ into a finite number of
bins. 

One proceeds as follows. The first step is to decide which of the
infinity of coexisting pairs of phases inside the cloud curve one would
like to determine for a given temperature $T$. To do so one specifies a
value for the parameter $\xi$ which measures the fractional volume of
the solid phase, and thus parameterizes the dilution lines between the
liquid phase cloud ($\xi=0$) and the solid phase cloud ($\xi=1$). The
general strategy for determining the coexistence properties at the given
$\xi$ (and $T$) is then to tune $p, \tilde\mu(\sigma)$  and $n\p$
iteratively, such as to simultaneously satisfy both a generalized lever
rule {\em and} equality of the a-priori probabilities of the phases,
ie.

\setcounter{abc}{1}
\bea 
\label{eq:methoda}
n\p\f\p(\sigma) &=& (1-\xi)\ronesig + \xi\rtwosig \:,\\
\addtocounter{abc}{1}
\addtocounter{equation}{-1}
{\mathcal{R}}_{\mbox{\sc {f,cs}}}&=&1\:.
\label{eq:methodb}
\eea 
\setcounter{abc}{0}
In the first of these constraints (Eq.~\ref{eq:methoda}), the ensemble
averaged daughter density distributions $\ronesig$ and $\rtwosig$ are
assigned by averaging only over configurations belonging to the fluid
and solid phases respectively. The deviation of the weighted sum of the
daughter distributions $\bar\rho(\sigma)\equiv(1-\xi)\ronesig +
\xi\rtwosig$ from the target $n\p\f\p(\sigma)$ is conveniently
quantified by a `cost' value:

\begin{equation}
\Delta\equiv\int \mid\bar\rho(\sigma)-n\p\f\p(\sigma)\mid d\sigma \;.
\label{eq:costfn}
\end{equation}
Requiring ${\mathcal{R}}_{\mbox{\sc {f,cs}}}=1$ in the second constraint
(Eq.~\ref{eq:methodb}), ensures that finite-size errors in coexistence
parameters are exponentially small in the system volume ~\cite{borgs1992,buzzacchi2006}.

The iterative determination of $p, \tilde\mu(\sigma)$ and $n\p$ such as
to satisfy Eqs.~\ref{eq:methoda} and \ref{eq:methodb} then proceeds thus:

\begin{enumerate}

\item Guess an initial value of the parent density $n\p$ corresponding
to the chosen value of $\xi$. (If one starts with a small degree of
polydispersity, a suitable estimate can be obtained from the  densities
of the coexisting phases in the known monodisperse limit, together with
the lever rule.)

\item Tune the pressure $p$ (within the HR scheme) such as to minimize $\Delta$.

\item Similarly tune $\tilde\mu(\sigma)$
(within the HR scheme, see below) such as to minimize $\Delta$.

\item Repeat steps 2 and 3 until $\Delta < {\rm Tolerance}$.

\item Measure the corresponding value of $\zratio$.

\item if $|\zratio-1|< {\rm Tolerance}$, finish, otherwise vary $n\p$ and repeat from step
2.

\end{enumerate}

The minimization of $\Delta$ with respect to variations in $p$ (step 2)
can be easily automated using standard 1-dimensional minimization algorithms such
as the ``brent'' routine described in Numerical Recipes
\cite{Numericalrecipes}. The same applies to the minimization of $|\zratio-1|$
with respect to variations in $n\p$ in step $6$. In step 3 the
minimization of $\Delta$ with respect to variations in
$\tilde\mu(\sigma)$ is most readily achieved \cite{wilding2002d} using
the following simple iterative scheme for $\tilde\mu(\sigma)$:

\begin{equation}
\tilde\mu_{k+1}(\sigma)=\tilde\mu_k(\sigma)+\alpha\ln\left( \frac{n\p f\p(\sigma)} {\bar\rho(\sigma)}\right)\;,
\label{eq:update}
\end{equation}
for iteration $k\to k+1$. This update is applied simultaneously to all
entries in the histogram of $\tilde\mu(\sigma)$, and thereafter the
distribution is shifted so that $\tilde\mu(\sigma_0)=0$. The quantity
$0<\alpha<1$ appearing in eq.~\ref{eq:update} is a damping factor, the
value of which may be tuned to optimize the rate of convergence. 

In practice we find that the minimization of $\Delta$ with respect to
variations in $\tilde\mu(\sigma)$ operates well provided
$\bar\rho(\sigma)$ is sufficiently close to $n\p
f\p(\sigma)$ for histogram reweighting to be effective. Note that (as
described in \cite{buzzacchi2006}) it is important that in steps $4$ and
$6$, one iterates to a very high tolerance in order to ensure that the
finite-size effects are exponentially small in the system size.
Typically we iterated until both $\Delta$ and $|\zratio-1|$ were less than
$10^{-12}$.

The values of $n\p$ and $p$ resulting from the application of the above
procedure are the desired parent density and pressure corresponding to the nominated
value of $\xi$. For $\xi=0$ and $\xi=1$, these are (respectively) the
liquid and solid cloud point densities and pressures, while shadow
points are given by the properties of the coexisting incipient daughter
phase, which may be simply read off from the appropriate peak positions
of the cloud point distributions of quantities such as the fluctuating
packing fraction. 

Finally in this section, we note that whilst in the interests of clarity
we have not included a description of temperature reweighting in our
procedure, it is straightforward to incorporate such. Doing so allows
one to scan the $p-T$ plane in a stepwise fashion, and thus, ultimately,
determine the entire transition region. 

\section{Solid-fluid coexistence of size-disperse soft spheres}

We have applied our methodology to obtain the cloud and shadow
point properties of a polydisperse system of softly repulsive spheres
described by the $r^{-12}$ potential

\begin{equation}
v(r)=\epsilon(\sigma_{ij}/r)^{12}\:,
\end{equation}
with $\sigma_{ij}=(\sigma_i+\sigma_j)/2$. In fact, for a given choice of
$f(\sigma)$ the thermodynamic state of this model does not depend on
$n\p$ and $T$ separately but only on the combination
$n\p(\epsilon/k_BT)^{1/4}$. Thus, the coexistence points for various
temperatures scale exactly onto one another \cite{Hoover1970} and one
therefore only needs to determine coexistence properties for a single
temperature. Accordingly in this work we shall consider the case
$(\epsilon/k_BT)=1$.

In our simulations, the interparticle potential was truncated at half
the box size and periodic boundary conditions were applied. Since for the
system sizes studied the value of the potential is extremely small at
the typical cutoff radius, no correction was applied for the truncation.

We chose to study a parent distribution of the top-hat form defined by

\begin{equation}
f(\sigma)=\left\{
\begin{array}{ll}
(2c)^{-1} & \mbox { if $1-c\le \sigma \le 1+c$} \\
~~0      &  \mbox { otherwise }
\end{array}
\right.,
\label{eq:th}
\end{equation}
where (without loss of generality), the mean particle diameter has been
set to $\bar{\sigma}=1.0$. In this initial study we report results for a
system of $N=256$ particles and rather narrow parent distributions
$c\leq 0.05$, for which the dimensionless degree of polydispersity
(ratio of standard deviation to mean of $f(\sigma)$ is
$\delta=c/\sqrt{3}\lesssim 3\%$.  For the purposes of forming
$\tilde\mu(\sigma)$ and $\rho(\sigma)$, each unit of $\sigma$ was
discretized into $500$ bins. 

As a preliminary step we determined the coexistence parameters of the
transition from fluid to face-centred-cubic (fcc) solid in the
monodisperse limit. This was done using the standard formulation of PSMC
(see sec.~\ref{sec:mono}) with the results \cite{Wilding2008}: 
$p_{coex}=22.32(3),\rho_F=1.148(9),\rho_{CS}=1.190(9)$. Thereafter we 
attempted to locate the fluid phase cloud point ($\xi=0$) for a narrow
top-hat parent having $c=0.01$. To this end we initialized the chemical
potential difference distribution as $\tilde\mu(\sigma)=0$ (for $0.99\le
\sigma \le 1.01$) and $\tilde\mu(\sigma)=-100$ otherwise, and assigned
the pressure the value $p=22.32$ pertaining to the monodisperse limit.
We then performed a long PSMC run, the results of which were reweighted
(using the procedure described in Sec.~\ref{sec:iterate} together with
the monodisperse value $n\p=1.148$ as the initial guess for the fluid
cloud point density), to yield accurate estimates of the fluid phase
cloud point pressure, parent density $n\p$ and chemical potential
difference distribution $\tilde\mu(\sigma)$, as well as the shadow phase
daughter distribution.

In order to progress to higher degrees of polydispersity, we proceeded
in a stepwise fashion. The form of $\tilde\mu(\sigma)$ previously
determined for $c=0.01$ was extrapolated via a quadratic fit to cover
the range $0.98\le \sigma \le 1.02$, while the pressure for $c=0.02$ was
estimated by linearly extrapolating the results for $c=0.00$ and
$c=0.01$. A new PSMC run was then performed, the results of which were
reweighted to give accurate estimates of the cloud point parameters for
the top hat parent having $c=0.02$.  In this manner we were able to
steadily increase the degree of polydispersity, measuring the cloud
point pressure and parent density as we went. In an identical
fashion we determined the dependence on polydispersity of the solid
cloud point parameters, by setting $\xi=1$ in the above procedure.

\begin{table}[h]
\begin{tabular}{c|cccc}
$c$    & $p_F$       & $n\p_F$       & $p_{CS}$      & $n\p_{CS}$\\ \hline
0    & 22.32(3)  & 1.148(9)   & 22.32(3)    & 1.190(9)          \\
0.02 & 22.46(3)    & 1.149(1)     & 22.49(3)   & 1.192(2)            \\
0.05 & 23.21(4)    & 1.159(1)     & 23.39(5)      & 1.201(2)             \\ \hline
\end{tabular}

\caption{Estimates of the coexistence pressure and parent density at the
fluid and fcc-solid cloud points for the half-widths $c$ of the
top-hat parent distribution indicated. }
\label{tab:clouddata}
\end{table}

\begin{figure}
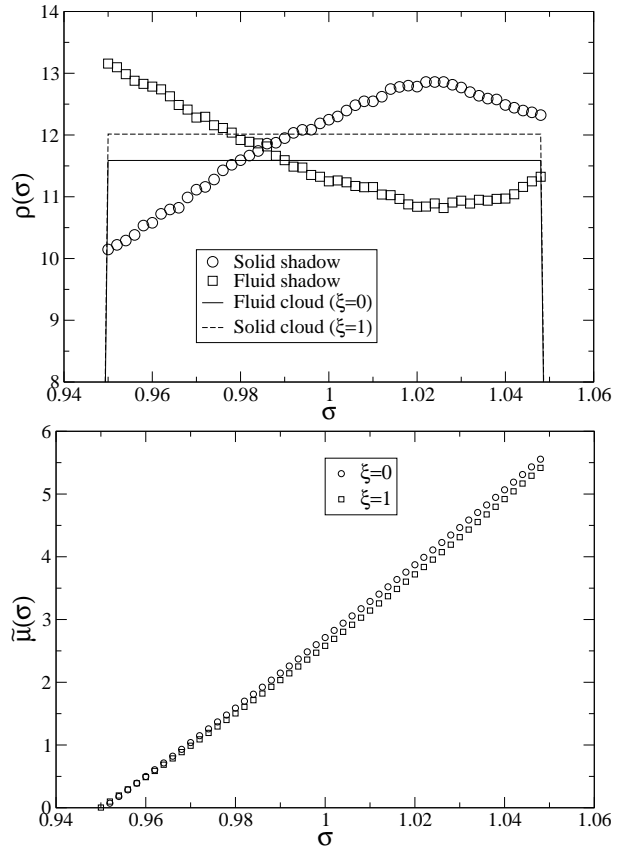

\vspace{5mm}\includegraphics[type=eps,ext=.eps,read=.eps,width=8.0cm,clip=true]{shadows}\\
\includegraphics[type=eps,ext=.eps,read=.eps,width=8.0cm,clip=true]{mudiff}\\
\caption{{\bf (a)} Estimates of the cloud and shadow point density distributions for
$c=0.05$. {\bf (b)} Corresponding distribution of chemical potential
differences, for $\sigma_0=0.95$. Statistical uncertainties do not exceed the symbol sizes.}
\label{fig:shadow}
\end{figure}

Table.~\ref{tab:clouddata} shows the fluid and solid cloud point
parameters for $c=0, 0.02$ and $0.05$. We note that our results for the
monodisperse limit ($c=0$) are consistent with existing estimates for
the same model obtained via thermodynamic integration \cite{Hoover1970}.
Our data for $c>0$ exhibit a clear separation of the cloud point
pressures as polydispersity increases, reflecting the
polydispersity-induced broadening of the coexistence region described in
Sec.~\ref{sec:intro}. Also apparent is an increase in the cloud point
densities with increasing polydispersity, a finding which is in accord
with theoretical predictions for polydisperse hard spheres
\cite{fasolo2004}. In Fig.~\ref{fig:shadow} we plot the form of the
shadow phase daughter distributions for $c=0.05$. This shows that at the
fluid phase cloud point the incipient solid phase contains a majority of
large particles compared to the parent, whilst at the solid phase cloud
point, the incipient (liquid) phase contains a majority of small
particles. This too is in accord with the predictions for models of
polydisperse hard spheres \cite{fasolo2004}. Fig.~\ref{fig:shadow}(b) 
shows the corresponding cloud point forms of the chemical potential
difference distribution $\tilde\mu(\sigma)$.

\begin{figure}
\vspace{5mm}
\includegraphics[type=eps,ext=.eps,read=.eps,width=8.0cm,clip=true]{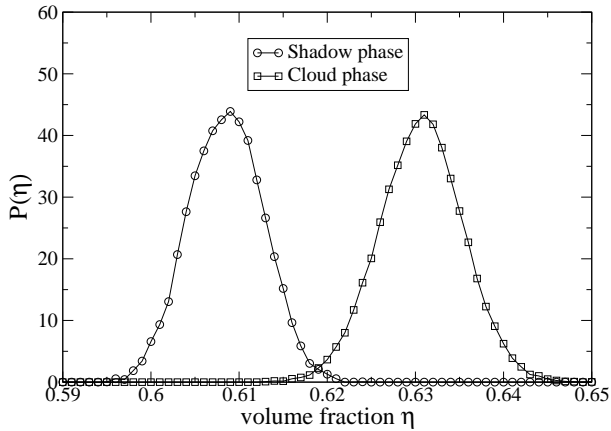}\\
\caption{The distribution of the fluctuating packing fraction
$P(\eta)$ in the solid cloud phase and in the shadow daughter phase for $c=0.05$, as described
in the text. Statistical uncertainties are comparable with the symbol sizes.}
\label{fig:peta}
\end{figure}

Finally, we consider how one obtains the properties of shadow phases
such as their average packing fraction or average number density. This
is illustrated in Fig.~\ref{fig:peta} which shows (for $c=0.05$) the
form of the probability distribution of the fluctuating packing fraction
at the solid cloud point ($\xi=1$), ie. the distributions
$P(\eta^{(1)})$ and $P(\eta^{(2)})$ with $\eta^{(i)}=(\pi/6)\int
\sigma^3\rho^{(i)}(\sigma)$. The peak at large $\eta$ corresponds to the
solid cloud phase and the peak at lower $\eta$ corresponds to the
fluid shadow phase. The average of the shadow point packing fraction
$\bar \eta$ can be simply read off from the position of the shadow
phase peak. For a system with a non-trivial temperature dependence of
the phase boundary, such plots (and analogous one for the number
density), obtained for a range of temperatures, allow construction of
the shadow curve in either the packing fraction or number density
representation.
 
\section{Conclusions}

In summary, we have presented an accurate approach for determining the
fluid-solid coexistence properties of polydisperse fluids under the
experimentally relevant constraint that the parent distribution has a
fixed functional form. The method is capable of determining the
properties of any of the infinity of coexistence state points that
pertain inside the cloud curve. By illustration we have determined the
cloud and shadow properties at the freezing transition for a system of soft
spheres having small degrees of polydispersity. 

Like the bare PSMC method on which it is based, our approach requires a
significant investment of computational effort in order to reap the
benefits of accurate estimates of coexistence properties. The results
presented here required about $4$ weeks of CPU time on an 8-core, 3GHz
processor. Nevertheless, it should be straightforward to extend our
study to higher degrees of polydispersity and larger system sizes,
albeit with a correspondingly increased computational expenditure. 

Thus our method could prove useful in helping to investigate interesting
theoretical  predictions \cite{fasolo2004} concerning the fluid-solid
coexistence properties of polydisperse fluids. Specifically, it has been
proposed that while on the fluid cloud curve the degree of
polydispersity of the fluid parent can be increased to rather large
values of $\delta \gtrsim 14\%$, in the coexisting (shadow) solid
daughter phase, $\delta$ never exceeds the more modest value
$\delta\approx 6\%$. On the other hand, on the solid phase cloud curve,
the single solid parent is itself predicted to become unstable at about
$\delta\approx 6\%$ and phase separates at a triple point into two solid
phases and a liquid. We hope to be able to address these issues in
future work.

\acknowledgments

The author thanks Peter Sollich for prompting his interest in this
problem and for a careful reading of the manuscript.

\bibliography{paper.bbl}
\bibliographystyle{prsty}

\end{document}